Partition functions of two-dimensional Ising models - A perspective from Gauss hypergeometric functions

M V Sangaranarayanan

Department of Chemistry Indian Institute of Technology- Madras Chennai-600036

India

E-mail: sangara@iitm.ac.in

Abstract

Employing heuristic susceptibility equations in conjunction with the well-known critical exponents, the magnetization and partition function for two-dimensional nearest neighbour Ising models are formulated in terms of the Gauss hypergeometric functions. The isomorphism existing between the Bragg-Williams approximation and Onsager's exact solution is pointed out. The precise manner in which the critical exponents influence the partition functions is pointed out for the first time.

1.Introduction

The exact solution of two-dimensional nearest neighbour Ising models has remained a Holy Grail in condensed matter physics, on account of its intractability of counting $2^N$ configurations for a square lattice of N sites when $N \to \infty$ [1]. Since the one-dimensional nearest neighbour Ising models do not predict the occurrence of the phase transition, the two-dimensional analogues have evoked considerable interest[2-4].

Onsager's exact solution of two-dimensional Ising models is a break-through in this context which yielded an explicit expression for zero-field magnetization *vis a vis* critical temperature [5]. Further insights regarding the Onsager's exact solution resulted from the analysis of Lee and Yang[6]. Among various approximations investigated in this context, mention should be made of the following: Bragg-Williams Approximation (BWA) [7], Bethe quasi-chemical approach [8], series expansions [9], renormalization group techniques, [10] scaling hypothesis [11], graph procedures [12] etc.

While the 'zero magnetic field' result is the most interesting one due to the occurrence of the phase transition, the non-availability of the partition functions and magnetizations for finite fields has been of serious concern in the context of binary alloys and absorbed systems where the chemical potentials are the counterparts of the magnetic fields [13]. Furthermore, the thermodynamic limit N $\to$

∞ is essential to predict the critical phenomena[14]. Nevertheless, extrapolation and series acceleration procedures employing ϵ-convergence algorithm [15], Pade' Approximants [16] can also yield accurate critical parameters using partial information. While rigorous procedures are essential for grasping the complexity of Ising models, any heuristic analysis is not unwelcome, since it can often provide fascinating insights-not obvious in the plethora of rigorous results. In this context, it is worth mentioning that (i) the precise role of the critical exponents in dictating the partition functions; (ii) correspondence between the Bragg-Williams approximation and Onsager's exact solution and (iii) new methods of re-formulating old equations have remained elusive.

The objectives of this Communication are to (i) deduce the magnetization and partition functions for the two-dimensional nearest neighbour Ising models using a heuristic susceptibility equation in conjunction with the universal critical exponents and (ii) demonstrate the isomorphism between the Bragg-Williams approximation and Onsager's exact solution from the perspective of Gauss hypergeometric functions.

2. Methodology

2.1 Critical exponents for the two-dimensional Ising model

Consider the two-dimensional nearest neighbor Ising Hamiltonian on a square lattice [3] viz

$$H_T = -J \left[ \sum_{\langle ij \rangle} (\sigma_i \sigma_j \sigma_i \sigma_{j+1} + \sigma_i \sigma_j \sigma_{i+1, \sigma j}) \right] - H \sum \sigma_{i,j} \qquad (1)$$

where $J$ denotes the interaction energy, H being the external magnetic field. The corresponding canonical partition function may be defined as

$$Q = Tr \{ e^{-\frac{H_T}{kT}} \} \qquad (2)$$

where $k$ is the Boltzmann constant. It is of interest to define various thermodynamic quantities using critical exponents:

zero-field magnetization: $M_{H=0} \sim |T_c - T|^\beta$  (3)

The isothermal magnetic susceptibility χ is defined as $\chi = (\partial M/\partial H)_T$ and its zero-field value (denoted as $\chi_{H=0}$) is given by the critical exponent γ

$\chi_{H=0} \sim |T-T_c|^{-\gamma}$  (4)

The field-dependence of the magnetization at the critical temperature $T_c$ in terms of the magnetic field H is as follows:

$|M|_{T=T_c} \sim |H|^{1/\delta}$  (5)

The eqn for the magnetization M as written here takes into account, the total number of sites N as well as the magnetic moment µ and the critical exponents satisfy the universality hypothesis [17]. The specific heat at H = 0 in terms of the critical exponent α is

$C_v$ (H=0) ~ |T-$T_c$|$^{-\alpha}$    (6)

2.2 Equation for the magnetic susceptibility

The preliminary step in this analysis involves formulating the equation for the isothermal magnetic susceptibility ( χ ) as

$$(kT)\chi = \frac{1 - M^{\frac{1}{\beta}}}{(\frac{T_c}{T})M^{(\delta-1)} - (\frac{T_c}{T} - 1)^{\beta(\delta-1)}} \quad (7)$$

using the critical exponents β and δ. Since the main focus of the study of the Ising models lies in the analysis of the critical properties, the above eqn is applicable for |T/$T_c$| ≤ 1. The rationale behind postulating the above eqn consists in recognizing that it leads to the correct critical exponents as

$$\chi (H = 0) \sim |(1 - \frac{T_c}{T})|^{-\beta(\delta-1)} \quad (8)$$

and

$$M(H = 0) = M_0 \sim |(1 - \frac{T_c}{T})|^{\beta} \quad (9)$$

It appears *prima facie* that since eqn (7) has been formulated incorporating the critical exponents, its validity may also be confined to T = $T_c$ . However, as demonstrated below, it is valid in the range T ≤ $T_c$ .

3. Two-dimensional Ising models under Bragg-Williams (mean field) approximation

The Bragg-Williams (or mean field) approximation constitutes zeroth order treatment in the analysis of Ising models and its prediction of the critical temperature in two dimensions is at variance with the exact value arising from Onsager's solution[5] . Nevertheless, it serves as a touch stone for verifying the validity of more improved versions such as Bethe approximation [8]. Consequently, the study of two-dimensional Ising modes using BWA is not devoid of significance and in what follows, the magnetization is analyzed with a view to obtain additional insights into Onsager's exact solution[5].

(A) Magnetization

The mean field critical exponents for the two-dimensional nearest neighbor Ising models under Bragg-Williams approximation have been deduced as β =1/2 and δ = 3 [4]. Incorporating the above values in eqn (7) yields the following eqn:

$$(kT)\chi = \frac{1-M^2}{\left(\frac{T_c}{T}\right)M^2 - \left(\frac{T_c}{T}-1\right)} \qquad (10)$$

Upon integrating the above, the magnetization is given by

$$M = \tanh\left(\frac{H}{kT} + M\frac{T_c}{T}\right) + \text{integration constant.} \qquad (11)$$

wherefrom the integration constant needs to be evaluated from the zero-field expression. Hence, if the exact eqn for the magnetization at H =0 is rendered available, the same for H ≠ 0 is obtainable using the susceptibility eqn. The zero-field magnetization $M_0$ under mean field approximation is given by [18]

$$M_0 = \tanh\left(M_0 \frac{T_c}{T}\right) \qquad (12)$$

and hence the integration constant in eqn (11) is zero. The foregoing analysis indicates that, commencing from the magnetic susceptibility eqn (7), it is possible to derive the field-dependence of the magnetization. This is not all. Integration of eqn (7) itself without substituting the mean field critical exponents leads to an interesting dependence of *M* upon *H* viz

$$\frac{H}{kT} = \frac{\left(\frac{T_c}{T}\right)M^\delta}{\delta} {}_1^2F\left(1,\beta\delta,\beta\delta+1,M^{\frac{1}{\beta}}\right) - \left(\frac{T_c}{T}-1\right)^{\beta(\delta-1)} M\, {}_1^2F\left(1,\beta,\beta+1,M^{\frac{1}{\beta}}\right) +$$
integration constant  (13)

where ${}_1^2F(a,b,c\,;\,z)$ denotes the Gauss hypergeometric function. If the integration constant is again assumed as zero and mean field values for the critical exponents β (= ½) and δ (=3) are substituted in the above, the following implicit eqn for the magnetization arises under mean field approximation *viz*

$$\frac{H}{kT} = \frac{\left(\frac{T_c}{T}\right)M^3}{3} {}_1^2F\left(1,\frac{3}{2},\frac{5}{2},M^2\right) - \left(\frac{T_c}{T}-1\right) M\, {}_1^2F\left(1,\frac{1}{2},\frac{3}{2},M^2\right) \qquad (14)$$

From the properties of the Gauss hypergeometric function [19], it follows that

$${}_1^2F\left(1,\frac{1}{2},\frac{3}{2},M^2\right) = (\tanh^{-1} M)/M \qquad (15)$$

and

$$\frac{{}_1^2F\left(1,\frac{3}{2},\frac{5}{2},M^2\right)}{3} = \frac{\left[{}_1^2F\left(1,\frac{1}{2},\frac{3}{2},M^2\right) - 1\right]}{M^2} \qquad (16)$$

Thus, the governing equation for magnetization becomes

$$M = \tanh\left(\frac{H}{kT} + M\frac{T_c}{T}\right)$$

and is identical with eqn (11) as anticipated. This formulation of the magnetization in terms of the Gauss hypergeometric functions is new but not entirely un-anticipated; however, it is reasonable to presume that a similar analysis will yield the exact eqn for the magnetization in the case of two-dimensional nearest neighbor Ising models. Since the Gauss hypergeometric functions traditionally arise from the solution of the generic Fuchsian differential equations [20], several contiguous relations apart from the nature of singularities may indeed be deciphered.

(B) Partition function

The conventional definition of magnetization in terms of the canonical partition function Q is [18]

$$M = \frac{1}{N}\left(\partial \log Q / \partial H\right)_T \qquad (17)$$

However, from eqn (7),

$$dH = \frac{\left(\frac{T_c}{T}\right)M^{\delta-1} - \left(\frac{T_c}{T} - 1\right)^{\beta(\delta-1)}}{1 - M^{\frac{1}{\beta}}} \, dM \qquad (18)$$

Upon substituting the above in eqn (17), the partition function is derived as

$$\frac{1}{N}\log Q = \frac{\left(\frac{T_c}{T}\right)M^{\delta+1}}{\delta + 1}{}_1^2F\left(1, \beta\delta + \beta, \beta\delta + \beta + 1, M^{\frac{1}{\beta}}\right)$$

$$- \frac{M^2}{2}\left(\frac{T_c}{T} - 1\right)^{\beta(\delta-1)} {}_1^2F\left(1, 2\beta, 2\beta + 1, M^{\frac{1}{\beta}}\right) + f1\left(M = 0, \frac{T}{T_c}\right) \qquad (19)$$

where $f1\left(M = 0, \frac{T}{T_c}\right)$ is an unknown function of the two variables indicated. By substituting the mean field critical exponents $viz$ $\beta$ (= ½) and $\delta$ (=3), the partition function becomes

$$\frac{1}{N}\log Q = \frac{\left(\frac{T_c}{T}\right)M^4}{4}{}_1^2F(1,2,3,M^2) - \frac{M^2}{2}\left(\frac{T_c}{T} - 1\right){}_1^2F(1,1,2,M^2) + f_1(M = 0, T/T_c) \qquad (20)$$

If the function $f_1(M=0, T/T_c)$ is identified as log (2), the above equation becomes identical with the partition function pertaining to the mean field approximation, as shown in Appendix A and log (2) corresponds to the value of $\frac{1}{N}\log Q$ at M =0. Thus, the mere formulation of the susceptibility eqn (7) exploiting the information on the critical exponents seems to be adequate for deducing the partition function, in general. Analogously, the known zero-field magnetization equations for any other models (such as Heisenberg, three-dimensional Ising etc) in conjunction with the critical exponents may enable the representation for magnetization when $H \neq 0$.

4. Onsager's exact solution for two-dimensional Ising models

The applicability of the susceptibility eqn (7) is anticipated even for the Onsager's exact analysis of two-dimensional Ising models *mutatis mutandis* if the critical exponents $\beta = 1/8$ and $\delta = 15$ are substituted. This aspect is investigated subsequently.

(A) Magnetization

Although the exact field-dependent magnetization for two-dimensional Ising models has not yet been deduced, the most important expression for M when H = 0 viz $M_0$ due to Onsager [5] is well-known *viz*

$$M_0 = (1 - \sinh^{-4}(2J/kT))^\beta \qquad (21)$$

with $J/kT_c = 0.4407$ for a square lattice. Further insights into the magnetization behavior near the critical region have been provided by Lee and Yang [6] according to which

$$M_0 \sim (1 - T/Tc)^{1/8} \qquad (22)$$

wherein the critical exponent $\beta = 1/8$. The other critical exponents $\gamma$ and $\delta$ have the values 7/4 and 15 respectively. It is also possible to estimate various critical exponents of Ising models using other methods such as ε- convergence algorithm [16], ratio method of Domb and Sykes [21], Pade' Approximants [22] etc. Upon substituting the exact values of the critical exponents $\beta = 1/8$ and $\delta = 15$ in eqn (7), it follows that

$$(kT)\chi = \frac{1 - M^8}{\left(\frac{Tc}{T}\right) M^{14} - \left(\frac{Tc}{T} - 1\right)^{\frac{7}{4}}} \qquad (23)$$

where $\chi = (\partial M/\partial H)_T$. Although the above eqn itself can be integrated to deduce the expression for M as a function of H and $T/T_c$, it is more illuminating if the general eqn (7) is employed wherefrom

$$\frac{H}{kT} = \frac{\left(\frac{Tc}{T}\right)M^{\delta}}{\delta} {}_1^2F\left(1,\beta\delta,\beta\delta+1,M^{\frac{1}{\beta}}\right) - \left(\frac{Tc}{T}-1\right)^{\beta(\delta-1)} M {}_1^2F\left(1,\beta,\beta+1,M^{\frac{1}{\beta}}\right)$$
$$+ integration\ const$$

as has been shown earlier in eqn (13). It is seen from the above that M ~ H $^{1/\delta}$ when T/T$_c$ =1. When H =0, the above eqn yields

$$\frac{\left(\frac{Tc}{T}\right)M_0^{14}}{15} {}_1^2F\left(1,\frac{15}{8},\frac{23}{8},M_0^8\right) = \left(\frac{Tc}{T}-1\right)^{\frac{7}{4}} {}_1^2F\left(1,\frac{1}{8},\frac{9}{8},M_0^8\right) + integration\ constant \quad (24)$$

if the exact values of β and δ are substituted. In contrast to the mean field approximation, the Gauss hypergeometric functions ${}_1^2F\left(1,\frac{1}{8},\frac{9}{8},M_0^8\right)$ and ${}_1^2F\left(1,\frac{15}{8},\frac{23}{8},M_0^8\right)$ do not seem to possess simple analytical expressions. The correctness of the above eqn can be inferred from the fact that when T/T$_c$ =1, the spontaneous magnetization vanishes. On the other hand, when T/T$_c$ = 0, the asymptotic behavior of the hypergeometric functions needs to be invoked.

(B) Partition function

The exact calculation of the partition function for two-dimensional Ising models when H ≠ 0 is 'almost impossible'. However, the extension of the earlier approach pertaining to the mean field approximation yields some new insights as shown below. The partition function in this case is also given by eqn (19) with β =1/8 and δ =15. Consequently,

$$\frac{1}{N}\log Q = \frac{\left(\frac{Tc}{T}\right)M^{16}}{16} {}_1^2F(1,2,3,M^8) - \frac{M^2}{2}\left(\frac{Tc}{T}-1\right)^{7/4} {}_1^2F\left(1,\frac{1}{4},\frac{5}{4},M^8\right) + f_2(M=0, T/Tc) \quad (25)$$

In contrast to the BWA, wherein the integration constant for (1/N) log Q equals log 2 at M = 0, this substitution is precluded for the exact analysis. Since the logarithmic singularity of the specific heat cannot be reproduced by a mere substitution of a constant term, the unknown f$_2$ (m=0, Tc/T) needs to be formulated such that Onsager's exact expression for the internal energy, specific heat etc arise at M = 0[23].

4.Perspectives

The present methodology enables the incorporation of the critical exponents explicitly into the partition function, in contrast to the hitherto-known procedures. In hindsight, the appearance of the Gauss hypergeometric function in the partition function is not unexpected since the Onsager's exact solution at H = 0 consists of the elliptic integral [24] which is related to the Gauss hypergeometric function [25]. The representation of the partition function in terms of more complicated Gauss hypergeometric functions is not entirely new and has recently been highlighted in case of two-dimensional Ising models for zero case [26]. Alternately, the random bond XY model can also be analysed using appropriate hypergeometric series [27].

It is of interest to enquire whether the partition functions from the mean field approximation and Onsager's exact analysis have a common structure. In order to respond to this, the identities involving Gauss hypergeometric functions may be invoked[28]. This leads to a re-formulation of eqn (25) as

$$\frac{1}{N}\log Q = -\left(\frac{T_c}{T}\right)\beta\left[\log\left(1-M^{\frac{1}{\beta}}\right)+M^{\frac{1}{\beta}}\right] + \beta\left(\frac{T_c}{T}-1\right)^{\beta(\delta-1)}\left[\log(1-M^2) + \sum_{1\leq l<1/2\beta} e^{-l\pi i/2}\log\{1-M^2 e^{l\pi i/2}\}\right] + f_2(M=0, T/T_c) \quad (26)$$

If β =1/2, δ =3, and f$_2$(M =0, T/Tc) = log (2), the above eqn is identical with the partition function for the mean field approximation. Hence the isomorphism between mean field approximation and Onsager's exact solution becomes transparent.

One of the limitations of the present investigation needs to be pointed out *viz* its non-validity in the case of the partition functions derived using Bethe approximation [8] on account of the incorporation of the short-range order parameter in addition to the magnetization (long range order parameter). However, the analysis is valid for Ising and Heisenberg models of diverse genre' since the only input parameters are the appropriate critical exponents and solutions (if known) for zero magnetic field.

It should be reiterated that extremely accurate polynomials for zero-field susceptibility as well as field-dependent magnetization have been rendered available using scaling arguments in conjunction with series expansions [29]. The formulation of the susceptibility eqn and subsequent analysis can yield elegantly, the partition function and magnetization when H ≠0. What demarcates the present study is the general framework underlying the partition function of Ising models apart from highlighting the explicit role of the critical exponents.

5. Summary

The equations pertaining to the partition function and magnetization for two-dimensional nearest neighbour Ising models are deduced using the susceptibility equations with appropriate critical exponents. The isomorphism between the Onsager's analysis and Bragg-Williams Approximation is pointed out from the perspective of Gauss hypergeometric functions.

Appendix A

In this Appendix, eqn(20) is rewritten using the well-known Gauss hypergeometric function identities in terms of elementary functions [25].

$$ {}_2^1F(1,2,3,M^2) = \frac{-2}{M^2} - \frac{2\log(1-M^2)}{M^4} \qquad (A1)$$

and

$$ {}_2^1F(1,1,2,M^2) = -\frac{\log(1-M^2)}{M^2} \qquad (A2)$$

Thus

$$ \frac{1}{N}\log Q = \frac{-\left(\frac{T_c}{T}\right)M^2}{2} - \frac{1}{2}\log(1-M^2) + \log 2 \qquad (A3)$$

which is identical with eqn (14.42) of [18]